\documentclass[printer]{aa}
\usepackage{psfig, wrapfig}

\begin{document}


\title{Dust-scattered X-ray halos around gamma-ray bursts:
GRB 031203 revisited and the new case of GRB 050713A\thanks{Based on
observations obtained with \emph{XMM-Newton}, an ESA science mission
with instruments and contributions directly funded by ESA Member
States and NASA} }

  \author{A. Tiengo\inst{1,2} and S. Mereghetti\inst{1}}

   \offprints{A. Tiengo, email: tiengo@mi.iasf.cnr.it}

  \institute{INAF - Istituto di Astrofisica Spaziale e Fisica Cosmica Milano,
          Via Bassini 15, I-20133 Milano, Italy
          \and
              Dipartimento di Fisica, Universit\`{a} degli Studi di Milano,
              via Celoria 16, I-20133 Milano, Italy
   }

  \date{}

\abstract{Scattering by interstellar dust grains can produce time
variable X-ray halos around gamma-ray bursts (GRB). In particular,
an X-ray expanding ring is expected when a short pulse of X-ray
radiation is scattered by a narrow layer of dust in our Galaxy. We
present a new method to detect and analyze dust scattering expanding
rings around gamma-ray bursts, using as an example \emph{XMM-Newton}
data of GRB 031203. Consistent with previous reports
(\cite{vaughan}), we find that the two  expanding rings observed in
this burst are due to dust unevenly distributed in two layers at
distances of 870 and 1384 pc, with the more distant one responsible
for 70\% of the total optical depth. Our modelling of the rings
indicates that the prompt X-ray emission of GRB 031203 was
consistent with a power law spectrum with photon index
$\Gamma$=2.1$\pm$0.2 and  1--2 keV fluence of
(3.6$\pm$0.2)$\times$10$^{-7}$ erg cm$^{-2}$. Thanks to the
sensitivity of our technique, we discovered an expanding ring around
another burst recently observed with \emph{XMM-Newton}, GRB 050713A.
In this case the dust layer is located at a distance of 364 pc and
we derive a GRB  fluence of (1.2$\pm$0.3)$\times$10$^{-7}$ erg
cm$^{-2}$  (1--2 keV). A search for similar halos in other twelve
bursts observed with \emph{XMM-Newton} gave negative results.
  }

\authorrunning{A. Tiengo \& S. Mereghetti}
\titlerunning{GRB halos}

\maketitle


\section{Introduction}

Diffuse X-ray halos, produced by scattering on interstellar dust
grains, have been observed around several bright sources (e.g.
\cite{MG86,PS95}). Since the scattering cross section depends on
the grain properties, the study of the energy-dependent radial
profile of the halos gives useful information on the dust grain
size, composition, and spatial distribution along the line of
sight (e.g. \cite{ML91,D03}). Most of the scattering halos
observed to date involved X-ray sources with a steady luminosity
(or with variations of amplitude and/or timescales not relevant
for the observable halo properties).

Halo photons scattered at larger radii suffer greater time delays
owing to their longer path lengths. Therefore, source variability
causes time-dependent changes of the halo radial profile. Knowing
the dust distribution along the line of sight it is possible to
derive the distance of variable sources by studying their halos
(\cite{ts73}). This method was suggested as a way to discriminate
between a galactic and a cosmological origin for Gamma-Ray Bursts
(GRBs, \cite{pac,klose}), but could not be applied to this problem
due to the lack of adequate X-ray telescopes.  Although the GRB
distance scale controversy has now been solved, the observation of
dust scattering halos around GRBs can provide other interesting
information and is within the capabilities of current satellites.
Recently, Draine \& Bond (2004) described how to measure the
distance of galaxies through the dust scattering halos of
background variable active galactic nuclei or GRBs.

The first detection of a GRB halo was obtained with a prompt
\emph{XMM-Newton} Target of Opportunity observation  of GRB 031203
(\cite{vaughan}), a burst at redshift z=0.1 quickly localized thanks
to the INTEGRAL Burst Alert System (\cite{ibas}). More recently
another dust scattering halo has been discovered around GRB 050724
using Swift (\cite{romano}).

Here we present a new method to detect and analyze  time variable
dust scattering halos like those expected for GRBs. As an
illustration of our technique we apply it to GRB 031203, using data
from the EPIC PN camera of \emph{XMM-Newton} (Sect.~2). Our results
on this burst provide an independent confirmation of those obtained
from the EPIC MOS data (\cite{vaughan}) for what concerns the dust
distance, but we derive a different estimate of the burst emission
properties (Sect. 3). We then analyze a recent \emph{XMM-Newton}
observation of another burst, GRB 050713A (Sect. 4), finding
evidence for the presence of a scattering halo due to a dust layer
located at a distance of 364 pc.

\begin{figure*}[ht!]
 \psfig{figure=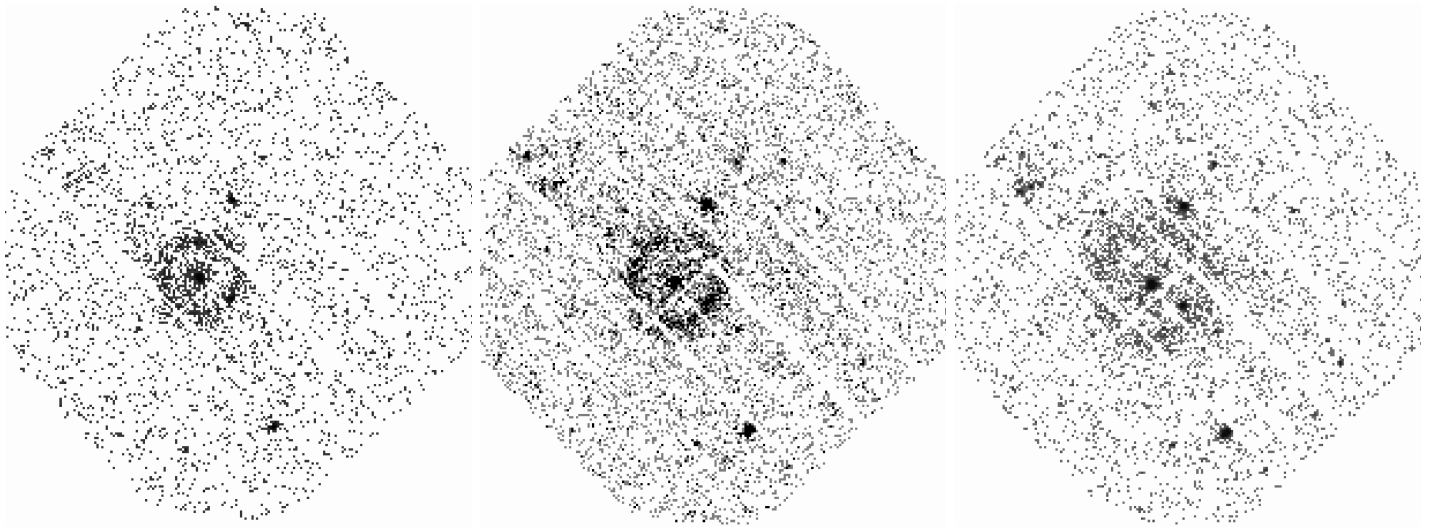}
\caption{EPIC PN images accumulated in three different time
intervals (23.5--30 ks, 30--45 ks, and 45--80 ks after the burst)
showing the dust scattering expanding rings around GRB 031203. The
images refer to the 1--2 keV energy range, where the halo spectrum
peaks (see Sect. 3). Each image covers a field of about
25$'\times$25$'$, North is to the top, East to the left. The PN
counts have been selected keeping the events with pattern 0--4 and
masking out the chip borders and defects. All the data have been
analyzed using SAS version 6.1.0. }
\end{figure*}

\section{Detection of  X-ray expanding rings}

The general expression for the time delay $t$ of photons from a
source at cosmological distance scattered by a dust layer at
redshift $z_d$ is given by

\smallskip

 $ t = (1+z_d) d_s d_d \Theta^2 / (2 c d_{ds})$,

\smallskip

\noindent  where $d_s$, $d_d$, and $d_{ds}$ are respectively the
angular diameter distances of the source, of the dust layer and
between the source and the dust layer, c is the velocity of light,
and $\Theta$ is the angle between the scattered photon and the
source direction (\cite{ms99}). When the scattering dust layer is in
our galaxy, at distance $d_s$, this equation simplifies to:

\smallskip

 $ t = (d_s/2c)~\Theta^2$.   ~~~~~~~~~~~~~~~~~~~~~~~~~~~~~~~~~~~~~~~~~~(1)

\smallskip

\noindent A short pulse of X-rays, emitted during a GRB, will
therefore produce a ring centered at the burst coordinates and with
radius $\Theta$ expanding with time according to eq. (1). The
intrinsic width $\Delta\Theta$ of the ring at a given time depends
in a similar way on the duration of the X-ray pulse and on the
thickness of the dust layer, while the observed width will also
depend on the angular resolution of the detector and on the
expansion of the ring during the exposure.

If the halo surface brightness is sufficiently high, the expanding
ring can be easily  detected by comparing X-ray  images taken at
different times, as illustrated in Fig.~1 for GRB 031203. The images
shown in Fig.~1 are based on data obtained with the PN camera
(\cite{pn}) of the \emph{XMM-Newton} EPIC instrument, during a
$\sim$56 ks long observation that started 6.5 hours after the burst.
We propose an alternative method to visualize and detect an
expanding X--ray ring, based on the construction of a
\textit{dynamical image} in which each count, detected with position
x$_i$,y$_i$ and arrival time T$_i$, is binned according to its new
coordinates
\smallskip

X$_i$ $\equiv$ T$_i$ -- T$_0$ = t$_i$

Y$_i$ $\equiv (x_{i}-x_{B})^{2} + (y_{i}-y_{B})^{2}$ =
$\Theta^2_i$

\smallskip
\noindent where T$_0$ is the time of the burst and x$_{B}$,
y$_{B}$ are its coordinates. Each point of a \textit{dynamical
image} corresponds to an angular distance from the GRB and to a
time delay. Celestial sources at fixed positions appear as
horizontal lines, while expanding rings centered at the GRB
position x$_{B}$, y$_{B}$ are visible as  inclined lines, with
angular coefficient inversely proportional to the scattering dust
distance d$_s$.

Fig.~2 shows the \textit{dynamical image} of GRB 031203, based on
exactly the same EPIC PN counts of Fig.~1. The two expanding
rings, produced by scattering in dust slabs at d$_s\sim$900 pc
and d$_s\sim$1400 pc (\cite{vaughan}) are clearly visible.

\begin{figure}[t]
 \psfig{figure=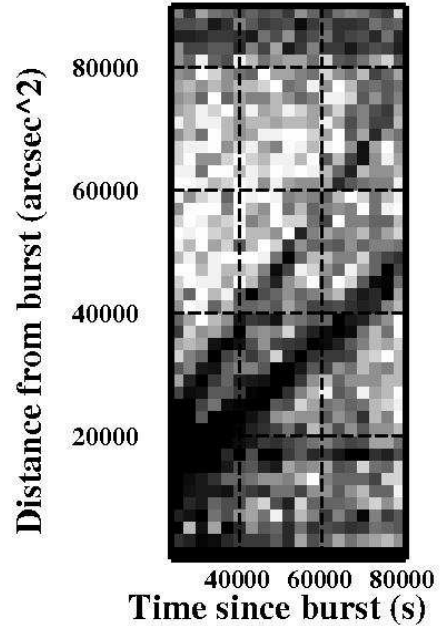,width=8.8cm}
\caption{Dynamical image of GRB 031203 based on the same EPIC PN
data as used in Fig.~1.}
\end{figure}

To estimate the dust distance and the flux in the halo we proceed
as follows. After removing the brightest point sources, for each
detected count we compute the quantity
\smallskip

$ D_i = 2c t_i / \Theta_i^2$   = 827  $t_i[s]~
\Theta_i^{-2}[arcsec] ~~~ \textrm{pc}$

\smallskip

\noindent The halo photons obey relation (1), contrary to the
background counts and photons from other sources in the instrument
field of view. Therefore, a halo is visible as a peak centered at
d$_s$ in the distribution n(D) of the D$_i$ values. A spatially
uniform instrumental background gives a contribution
proportional to D$^{-2}$ in the range from D$_{min}$ = 2c
t$_{max}$/$\Theta^2_{max}$ to  D$_{max}$ = 2c
t$_{min}$/$\Theta^2_{min}$, where t$_{min}$, t$_{max}$,
$\Theta_{min}$ and $\Theta_{max}$ delimit the rectangular region
of the dynamical image from which the counts are extracted.

Fig.~3 shows the distribution of D$_i$  (in the range
D$_{min}$--D$_{max}$) obtained selecting from the dynamical image
of GRB 031203 the events with 1$'<\Theta<10'$. The two peaks
corresponding to the dust scattering rings are clearly visible,
superimposed on the background contribution. A good fit to the
data in Fig.~3 ($\chi_{red}^2$=1.00 for 143 d.o.f.) is obtained
with the sum of a power law with index $\alpha=-1.77\pm$0.02 and
two Lorentzian curves centered at 870$\pm$5 pc and 1384$\pm$9 pc,
and with FWHM of 82$^{+17}_{-14}$ pc and 240$\pm$30 pc,
respectively (all the errors are at the 90\% c.l. for a single
parameter of interest). These widths are compatible with the
expected values of $\sim$100 pc and $\sim$200 pc, respectively,
caused by the instrumental Point Spread Function of 6$''$. The net
number of halo counts, obtained by integrating the two
Lorentzians, are 840$^{+210}_{-180}$ counts for the outer ring
(d$_s$=870 pc) and 1740$^{+270}_{-240}$ counts for the inner ring
(d$_s$=1384 pc).

\begin{figure}[t]
 \psfig{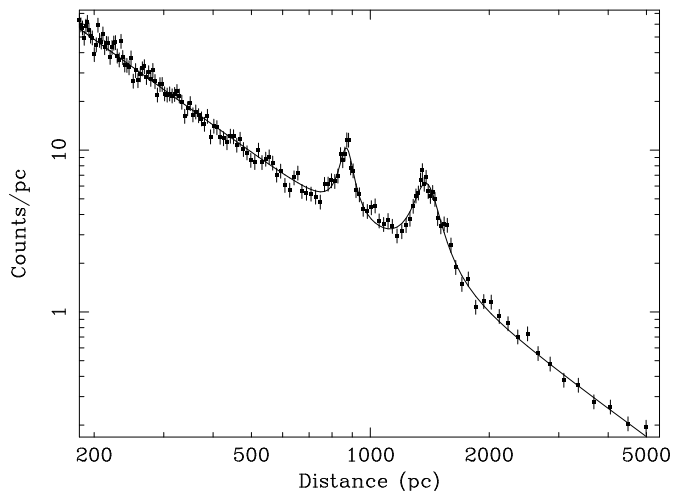}
\caption{Distribution  n(D) (see Sect. 2) for the EPIC PN
observation of GRB 031203. The data are binned so that each bin
contains 100 counts. The model is a power-law of index --1.77 plus
two Lorentzians centered at 870 pc and 1384 pc.}
\end{figure}

\section{Spectral analysis and prompt X-ray emission of GRB 031203}

When both the scattered and unscattered radiation are measured, it
is possible to deduce the properties of the dust. On the other
hand, for GRB halos, we usually can measure only the scattered
component of an X-ray pulse of unknown intensity emitted at (or
close to) the time of the burst. In order to derive some
information on such prompt X-ray emission from the observed halo,
we must assume the dust properties and the optical depth for X-ray
scattering (or deduce them from independent measurements). In the
following, we adopt a model of dust composed by carbonaceous and
silicate grains which is in good accord with several observations
(\cite{D03}) and we assume A$_V$=2 for the total extinction due to
the two dust layers in the direction to GRB 031203 (\cite{nk}).

The fluence F$_H$ of the radiation scattered in the halo between
the angles $\Theta_1$ and $\Theta_2$ is related to that emitted
from the source F(E) by the following relation:

\smallskip

F$_H$(E,$\Theta_1$,$\Theta_2$) = F(E) $\tau(E)$ (g($\Theta_2$,E)
-- g($\Theta_1$,E)) ~~~~~(2)

\smallskip

\noindent where g($\Theta$,E) is the fraction of halo photons
scattered at angles smaller than $\Theta$ and $\tau(E)$ is the
scattering optical depth of the dust layer, related to the optical
absorption A$_V$ by  $\tau\approx$ 0.15 A$_V$ E$^{-1.8}$(keV)
(\cite{DB04}). For a narrow scattering layer close to the
observer, which applies to our case, the halo profile is well
approximated by

\smallskip

g($\Theta$,E) $\approx$ ($\Theta/\Theta_s$)$^2$ /
(1+($\Theta/\Theta_s$)$^2$) ~~~~~~~~~~~~~~~~~(3)

\smallskip

\noindent with $\Theta_s$ = 360$''$/E(keV) (\cite{D03}). Fitting
the halo spectrum with eq. (2) it is then possible to derive the
spectrum and intensity of the prompt emission.

To extract the background subtracted count spectra of the two rings
we have applied the procedure described in Section 2 to the
distributions n(D) relative to 16 energy channels in the range
0.8-2.2 keV (the halo has a very soft spectrum and is not visible at
higher energy. Due to the small number of counts in each
distribution, we have kept the centers and widths of the Lorentzians
fixed at the best fit values previously found for the total energy
range. We described F(E) as an absorbed power law with photon index
$\Gamma$, and implemented equation (2) as a spectral model in the
XSPEC fitting package. This was used to fit simultaneously the
counts spectra of the two rings. To reduce the number of free
parameters, we kept the interstellar absorption fixed at the value
derived from the afterglow emission N$_H$=8.8$\times$10$^{21}$
cm$^{-2}$ (\cite{vaughan}) and linked the optical depths of the two
dust layers using the total value corresponding to A$_V$=2:
$\tau_{870 pc}+\tau_{1384 pc}$= 0.3 E$^{-1.8}$.

The best fit yielded a photon index $\Gamma$=2.1$\pm$0.2 and a
fluence of (3.6$\pm$0.2)$\times$10$^{-7}$ erg cm$^{-2}$ (1--2 keV)
for the GRB emission, and A$_V^{870 pc}$=0.6$\pm$0.05
(corresponding to A$_V^{1384 pc}=2-$A$_V^{870 pc}$=1.4$\pm$0.05).

These values can be compared with those obtained with a different
method and using the EPIC MOS data by Vaughan et al. (2004). They
concluded that, to match the power law photon index of the halo
(3.3$\pm$0.15, from a spectrum extracted only in the first part of
the observation), a value of  $\Gamma\sim$2 is required for the
prompt X-ray emission. They also found that the best fit to the
structure of the two rings is obtained by attributing 2/3 of the
total column density to the more distant dust slab. These results
are in accord with those derived in our analysis. Vaughan et al.
obtain a time integrated flux for the prompt emission of
1600$\pm$800 photons cm$^{-2}$ keV$^{-1}$ at 1 keV, while the
revised value of 1320$\pm$260 photons cm$^{-2}$ keV$^{-1}$ at 1 keV,
has been recently reported by the same group (\cite{wat}). For a
power law spectrum with $\Gamma$=2, the latter value corresponds to
a fluence of (1.5$\pm$0.3)$\times10^{-6}$ erg cm$^{-2}$ in the 1-2
keV range, a factor $\sim$4 higher than our measurement. This
discrepancy can be in part attributed to the different relation
between $\tau$ and A$_V$ adopted by these authors.

\section{The halo of GRB 050713A}

The advantages of using the dynamical images are particularly
evident when the  scattering halos are not easily visible in the
normal images, as in the case described here for GRB 050713A.

GRB 050713A was discovered and promptly localized (\cite{050713d})
with the Swift satellite, which also detected its X-ray afterglow
(\cite{050713xrt}). A Target of Opportunity observation was carried
out with \emph{XMM-Newton} (\cite{050713x,deluca}) starting $\sim$23
ks after the GRB prompt emission. Part of the $\sim$30 ks long
observation was badly affected by soft protons flares. Filtering out
the time intervals with the highest background reduces the EPIC PN
net exposure time to only 9.2 ks. The corresponding image in the
0.5--2 keV range (Fig.~4) shows the presence of some diffuse
emission around the GRB position, but the short exposure time does
not allow us to extract time resolved images sensitive enough to
detect an expanding ring.

\begin{figure}[t]
 \psfig{figure=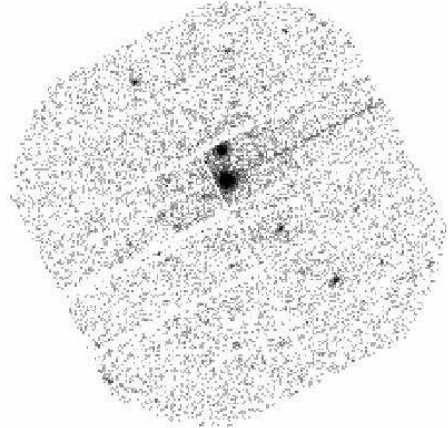,width=8.8cm}
\caption{EPIC PN image of GRB 050713A in the 0.5--2 keV energy
range. North is to the top, East to the left. The time intervals of
highest background have been excluded, giving a net exposure time of
9.2 ks. The GRB afterglow is the southern of the two brightest
sources.}
\end{figure}

\begin{figure}[t]
  \psfig{figure=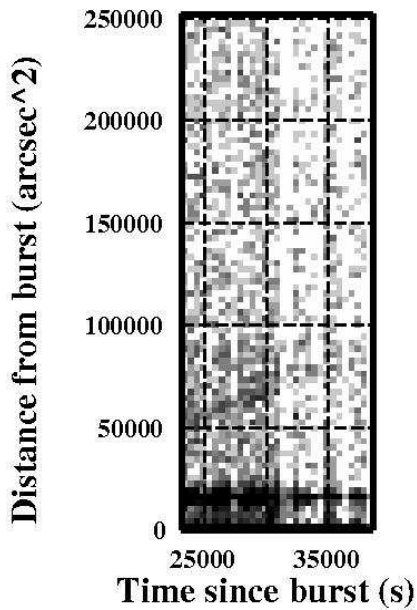,width=8.8cm}
\caption{Dynamical image of GRB 050713A. It was extracted using
exactly the same selections as used in Fig.~4. The bright
horizontal line is the source $\sim$2 arcmin North of the GRB,
while the faint oblique segment is the expanding ring.}
\end{figure}

On the other hand, the presence of an expanding halo is manifested
by  the inclined line starting at $\Theta^2\sim$50,000 arcsec$^2$
visible in the dynamical image (Fig.~5). More quantitative
information is obtained from the distribution n(D) shown in
Fig.~6.  This has been obtained with the counts in the 0.5-2 keV
energy range and excluding the brightest point sources. A fit with
a power-law gives a $\chi_{red}^2$ of 1.70 for 92 d.o.f., while
the addition of a Lorentzian centered at 364$^{+6}_{-7}$ pc
improves it to 1.23 for 89 d.o.f.. The Lorentzian FWHM is
33$^{+18}_{-12}$ pc, consistent with the width of 30 pc caused by
the PN angular resolution, and the total number of counts in the
Lorentzian is 185$^{+120}_{-90}$. The best fit power-law index is
--1.65$\pm$0.03.

\begin{figure}[t]
 \psfig{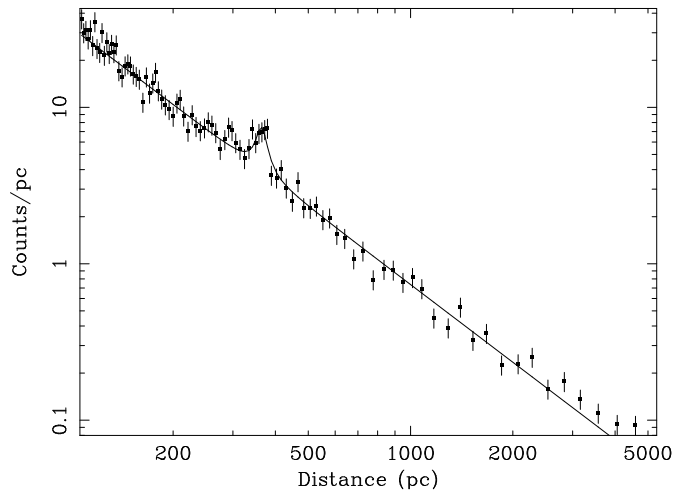}
 \caption{Distribution of n(D) (see Sect.4) for the EPIC PN
observation of GRB 050713A. The data are binned so that each bin
contains 50 counts. The model is a power-law of index --1.65 plus
a Lorentzian centered at 364 pc.}
\end{figure}

Since in this case the presence of the peak is less compelling
than in GRB 031203, we have also analyzed the MOS data. The MOS
exposure started $\sim$40 minutes before the PN one and the MOS
detectors are less sensitive than the PN to soft protons;
therefore, after the exclusion of the time intervals with the
highest background, the MOS exposure time is $\sim$2 times larger
than the PN one. Figure 7 shows the distribution n(D) (in the
range D$_{min}$--D$_{max}$ corresponding to 1$'<\Theta<$10$'$)
obtained from the events of the two MOS units selected in the
0.5--2 keV energy range, with pattern 0--12, and excluding the
point sources. The fit results ($\chi_{red}^2$=1.17 for 62 d.o.f.)
are consistent with those obtained with the PN data: a power-law
with index $\alpha$=$-1.67\pm$0.02 and a Lorentzian centered at
364$\pm$6 pc with a FWHM of 45$^{+26}_{-16}$ pc. The total number
of counts in the Lorentzian is 290$^{+200}_{-140}$, consistent
with the value found with the PN, that has a $\sim$30\% larger
effective area than the sum of the two MOS but a factor 2 shorter
exposure time.

Due to the low number of counts detected in the halo of GRB 050713A,
we used only three energy channels (0.5--1 keV, 1--1.5 keV, and
1.5--2 keV) to extract the halo spectra. The PN and MOS spectra were
fit simultaneously using the same model applied to GRB 031203, with
an appropriate correction factor to account for the time gaps
produced by the soft protons filtering and for the different
exposure times of the two instruments\footnote{the presence of time
gaps translates into an incomplete sampling of the halo radial
profile, that can be accounted for by a generalization of eq.(2)}.
We assumed A$_V$=0.5 mag (\cite{hak}) and the X--ray absorption
measured for the afterglow, N$_H$=3.25$\times$10$^{21}$ cm$^{-2}$
(\cite{deluca}). The best fit yielded a power law photon index
$\Gamma$=1.4$\pm$0.6 and a fluence of (1.2$\pm$0.3)$\times$10$^{-7}$
erg cm$^{-2}$ (1--2 keV unabsorbed) for the prompt GRB emission.

The 1-2 keV fluence obtained by extrapolating the spectrum measured
in the 15-350 keV range by Swift-BAT  (\cite{050713bat}) is
(3.6$\pm$1.1)$\times10^{-7}$ erg cm$^{-2}$, slightly higher than the
value derived from our analysis.

\begin{figure}[t]
 \psfig{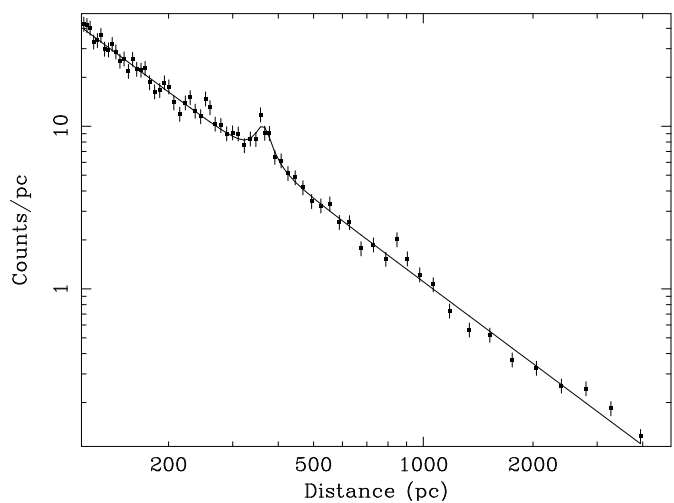}
 \caption{Distribution  n(D)  for the EPIC
 MOS observation of GRB 050713A. The data are binned so that each bin
contains 50 counts. The model is a power-law of index --1.67 plus
a Lorentzian centered at 364 pc.}
\end{figure}

\section{Search for halos around other GRBs}

To search for other dust scattering halos,  we analyzed the EPIC PN
data of all the  GRB observations performed after a delay smaller
than one day. The sample consists of the 14 GRBs listed in Table 1.
Except for the two cases discussed above, the n(D) distributions did
not show the presence of well defined peaks. For eleven GRBs, good
fits ($\chi_{red}^2$ in the range 0.9-1.2) were obtained with power
laws indexes $\alpha$ in the range from --2 to --1.9 (the results of
the fits are summarized in Table 1). This range of $\alpha$ values
is consistent with that ($-2.07$ to $-1.87$) found by analyzing the
n(D) distributions of several EPIC PN observations of blank fields
(we assumed $x_B$ and $y_B$ at the center of the field of view and
T$_0$ at 20,000 s before the start of the observation).

Only for GRB 050730 a power law gave a bad fit, but no well
defined peaks are present in its n(D) distribution. A careful
inspection of the PN image shows the presence of some diffuse
emission in the field of GRB 050730, which is likely responsible
for the  deviations from the distribution expected for a uniform
background.

\begin{center}
\begin{table*}
\caption{GRBs observed by \emph{XMM-Newton} within one day from the
burst.}
    \begin{tabular}[c]{lrrccccc}
\hline \hline
GRB & l & b & N$_{\rm H}$ & t$_{\rm start}$  & t$_{\rm stop}$ & $\alpha$ & $\chi^2_{red}$(d.o.f.)\\
name & ($^{\circ}$) & ($^{\circ}$) & (10$^{20}$ cm$^{-2}$) & (s) & (s) & & \\
\hline \hline
GRB010220  & 135 & 1 & 86 & 63370 & 97200 & --1.96$\pm$0.01 & 1.11(286) \\
GRB011211  & 275 & 36 & 4.2 & 43630 & 73410 & --1.92$\pm$0.01  & 0.92(168) \\
GRB020321  & 308 & --23 & 8.2 & 42930 & 88110 & --1.958$\pm$0.01 & 1.21(250) \\
GRB020322  & 113 & 29 & 4.6 & 55650  & 81950 & --1.967$\pm$0.009  & 1.08(449) \\
GRB030227  & 181 & --14 & 22  & 46740  & 77110 & --1.95$\pm$0.01  & 1.19(181) \\

GRB031203  & 256 & --5 & 60 & 23500  & 79830 & --1.65$\pm$0.02  & 7.6(149) \\

GRB040106  & 293 & 15 & 8.6 & 22100 & 63160 & --1.92$\pm$0.02  & 1.06(174) \\
GRB040223  & 342 & 3 & 60 & 24100  & 60030 & --1.91$\pm$0.02  & 0.89(128) \\
GRB040827  & 346 & 34 & 8.4 & 25240  & 76580 & --1.95$\pm$0.01  & 0.94(222) \\
GRB050223  & 331 & --19 & 7.1 & 38830  & 96940 & --1.975$\pm$0.008  & 1.00(601) \\
GRB050326  & 306 & --46 & 4.5 & 35530  & 75850 & --1.93$\pm$0.03  & 1.16(50) \\

GRB050713A  & 112 & 19 & 11 & 23160  & 46260 & --1.65$\pm$0.03  & 1.70(92) \\

GRB050730  & 337 & 54 & 3.1 & 29520  & 59850 & --1.74$\pm$0.02  & 2.27(109) \\
GRB050820B  & 288 & --17 & 8.4 & 21360  & 78000 & --1.962$\pm$0.008  & 1.08(734) \\
\hline \hline

\end{tabular}

\footnotesize{The Table reports the GRB Galactic coordinates, the
Galactic absorption (\cite{dl}), the start and stop of the
observation with respect to the time of the GRB prompt emission,
the power law slope fitted to the n(D) distribution, and the
$\chi^2$ value and number of degrees of freedom of the fit. }

\end{table*}
\end{center}

\section{Discussion}

In principle, from the observation and modelling of a dust
scattering halo, it is possible to obtain useful information on the
prompt GRB X--ray emission and the early phases of the afterglow,
even when these are not directly observed. This is particularly
interesting in the case of GRB 031203, since it has been suggested
that this low red-shift burst (z=0.105, \cite{pro}) was unusually
under-energetic, despite having the properties of a normal, long
duration GRB (\cite{sazonov}). The high value of the soft X--ray
fluence inferred from the dust halo is relevant for the overall
energetics and spectral shape of this burst.

In the 20-200 keV energy range observed with INTEGRAL, GRB 031203
had a fluence of (2.0$\pm$0.4)$\times$10$^{-6}$ erg cm$^{-2}$, a
spectrum  well described by a power law with photon index
$\Gamma$=1.63$\pm$0.06, and a duration of about 40 s
(\cite{sazonov}). Watson et al. (2005) derived an X-ray fluence
about one order of magnitude above the extrapolation of the INTEGRAL
spectrum. To explain such a large discrepancy, these authors
concluded that GRB 031203 consisted of two distinct pulses: the
first one with a hard spectrum detected by INTEGRAL, followed by a
softer one responsible for the bright X--ray scattering halo. The
X-ray fluence obtained with our method, although smaller than that
found by Watson et al., still indicates that the X-ray emission from
GRB 031203 was not a simple extrapolation of that measured above 20
keV. This is also consistent with the fact that the spectrum we
obtained for E$<$2 keV  ($\Gamma$=2.1$\pm$0.2), is steeper than that
measured with INTEGRAL ($\Gamma$=1.63$\pm$0.06, \cite{sazonov}).

We have discovered that also GRB 050713A has a dust scattering halo,
produced by a thin ($<$50 pc) dust layer at a distance of 364$\pm$7
pc. This distance is small enough to be compatible with the rather
high Galactic latitude (b=18.8$^{\circ}$) of this burst.
Extrapolating to the 1--2 keV energy range the spectrum measured
above 15 keV with Swift-BAT (a power law with
$\Gamma$=1.58$\pm$0.07, \cite{050713bat}), gives a fluence a factor
3$^{+2.2}_{-1.3}$ higher than that we estimated from the analysis of
the scattering halo. Contrary to the case of GRB 031203, the
difference could be explained by a spectral steepening between the
\emph{XMM-Newton} and Swift-BAT energy range. However, the spectral
slope we inferred for the X--ray emission, although poorly
constrained ($\Gamma$=1.4$\pm$0.6), is consistent with that
measured with Swift-BAT. Therefore, another possibility to reconcile
the fluence values is that we overestimated the amount of dust in
the scattering layer, which was derived from the assumption of
A$_V$=0.5 mag.

Finally, we note that the best fit values of the power-law indexes
$\alpha$ found for the n(D) distributions of GRB 031203 and GRB
050713A are significantly different from those seen in
\emph{XMM-Newton} blank fields and in observations of other GRB
afterglows. The latter have values of $\alpha$ only slightly flatter
than --2, as expected due to the non--uniform background
distribution over the field of view. The large values of
$\alpha$=--1.8 and --1.7 seen only in GRB 031203 and 050713A could
be somehow related to the presence of the scattering halos.
Different phenomena might produce this effect. Two equivalent
possibilities are scattering of the prompt X--ray emission by dust
distributed at different distances along the line of sight, and the
scattering of the X--ray afterglow (i.e. emission occurring after
the GRB) by the same dust layers responsible of the Lorentzian
peaks. Another likely effect is the presence of photons undergoing
more than a single scattering by the dust grains. Note that in this
case there is no univocal relation between time delay $t$ and
$\Theta$. The resulting faint diffuse emission around the GRB
position  will only affect the continuum distribution of n(D),
leaving the peaks due to single scattering at definite distances
unaltered.

\section{Conclusions}

We have developed  a simple method to  analyze dust scattering
expanding rings around GRBs, and applied it to \emph{XMM-Newton}
data of GRB 031203, the first burst to show a time variable dust
scattering halo (\cite{vaughan}). Similar to these authors, who used
a different method and data set,  we found that the dust responsible
for the two observed expanding rings is unevenly distributed in two
narrow ($<$100 pc and $<$270 pc, respectively) layers at distances
of 870$\pm$5 and 1384$\pm$9 pc, with the more distant one
responsible for $\sim$70\% of the total optical depth.

Thanks to the sensitivity of our method, we could show that also GRB
050713A has a dust scattering halo, produced by a thin ($<$50 pc)
dust layer at a distance of 364$\pm$7 pc. Moreover this method
allows very accurate determinations of the dust cloud distances.

We can expect that a growing number of dust scattering halos
around GRBs will be detected in the near future, thanks to the
rapid GRB localizations provided by currently operating satellites
and the possibility to perform sensitive follow-ups with good
quality X-ray imaging instruments. Most of these halos will
probably be relatively faint and not easily visible in time
resolved images,  while  using the method proposed here it will be
possible to detect them.

\emph{XMM-Newton} observations of mid galactic latitude GRBs are
ideal to uncover more dust scattering halos, thanks to the high
sensitivity and large field of view of the EPIC instrument. The
response time for \emph{XMM-Newton} rapid follow-ups (from $\sim$5
to 8 hours) is well matched to the time delays expected from dust
layers in our Galaxy at distances in the range from $\sim$100 pc to
a few kpc, and the angular resolution of EPIC is adequate for the
resulting structures, typically of the order of a few arcmin radius.
Earlier observations could reveal rings with smaller radii, but this
also requires a good angular resolution, as obtainable e.g. with
Chandra\footnote{which, however, has in general a longer reaction
time than \emph{XMM-Newton} for GRB follow-up observations}, to
disentangle the emission of the expanding halo from that of the
central afterglow source.

It is evident from the two  GRB halos discussed above that the
derivation of quantitative information on the unscattered X-ray
emission is presently subject to large systematic uncertainties due
to the assumptions required in the halo modelling. In this respect,
the observation  with \emph{XMM-Newton} and/or Swift-XRT, of dust
halos around GRBs for which bright prompt X-ray (or early afterglow)
emission is detected by Swift would be extremely useful, since it
could give the chance to compare the X--ray properties inferred from
the halo with those actually observed. The dust halo reported around
GRB 050724 (\cite{romano}) has already demonstrated the Swift
capabilities to detect such features. A systematic application of
the method presented here to all GRB follow-up observations would
possibly enlarge the sample through the detection of fainter halos,
not immediately seen in the traditional X--ray images.

\begin{acknowledgements}
This research has been partially supported by the Italian Space
Agency. We thank N.Schartel and the staff of the \emph{XMM-Newton}
Science Operation Center for performing the GRB follow-up
observations.
\end{acknowledgements}


\begin{thebibliography}{}

\bibitem[De Luca et al. 2005]{deluca} De Luca A. et al. 2005, GCN Circ. n. 3695
\bibitem[Dickey \& Lockman 1990]{dl} Dickey J.M. \& Lockman F.J. 1990, ARA\&A 28, 215
\bibitem[Draine 2003]{D03} Draine B.J. 2003, ApJ 598, 1096
\bibitem[Draine \& Bond 2004]{DB04} Draine B.J. \& Bond N.A. 2004, ApJ 617, 987
\bibitem[Falcone et al. 2005]{050713d} Falcone A. et al. 2005, GCN Circ n. 3581
\bibitem[Hakkila et al. 1997]{hak} Hakkila, J., Myers, J. M., Stidham, B. J., \& Hartmann, D. H. 1997, AJ 114,2043
\bibitem[Loiseau et al. 2005]{050713x} Loiseau N. et al. 2005, GCN Circ. n. 3594
\bibitem[Mathis \& Lee 1991]{ML91} Mathis J.S. \& Lee C.-W. 1991, ApJ 376, 490
\bibitem[Mauche \& Gorenstein 1986]{MG86} Mauche C.W. \& Gorenstein P. 1986, ApJ 302, 371
\bibitem[Mereghetti et al. 2003]{ibas} Mereghetti S.,  G\"{o}tz D., Borkowski J., Walter R. \& Pedersen H., 2003, A\&A 411, L291
\bibitem[Miralda-Escud\'{e} 1999]{ms99} Miralda-Escud\'{e} J. 1999, ApJ 512, 21
\bibitem[Morris et al. 2005]{050713xrt} Morris D. et al. 2005, GCN Circ. n. 3606
\bibitem[Neckel \& Klare 1980]{nk} Neckel T. \& Klare G. 1980, A\&AS 42, 251
\bibitem[Klose 1994]{klose} Klose S. 1994, A\&A 289, L1
\bibitem[Palmer et al. 2005]{050713bat} Palmer D. et al. 2005, GCN Circ. n. 3597
\bibitem[Paczynski 1991]{pac} Paczynski B. 1991, Acta Astronomica 41, 257
\bibitem[Predehl \& Schmitt 1995]{PS95} Predehl P. \& Schmitt J.H.M.M. 1995, A\&A 293, 889
\bibitem[Prochaska et al. 2004]{pro} Prochaska J.X. et al. 2004, ApJ 611, 200
\bibitem[Romano et al. 2005]{romano}Romano P. et al. 2005, GCN Circ. n. 3685
\bibitem[Sazonov et al. 2004]{sazonov} Sazonov S.Y., Lutovinov A.A. \& Sunyaev R.A. 2004, Nature 430, 646
\bibitem[Str{\"u}der   et al. 2001]{pn} Str{\"u}der L. et al. 2001, A\&A 365, L18
\bibitem[Tr\"{u}mper \& Sch\"{o}nfelder  1973]{ts73}Tr\"{u}mper J. \& Sch\"{o}nfelder V. 1973, A\&A 25, 445
\bibitem[Vaughan et al. 2004]{vaughan} Vaughan S., Willingale R., O'Brien P.T., et al. 2004, ApJ 603, L5
\bibitem[Watson et al. 2005]{wat} Watson D., Vaughan S.A., Willingale R., et al. 2005, ApJ Letters in press, astro-ph/0509477

\end{thebibliography}
\end{document}